\documentclass{iopart}
\usepackage{amsfonts,amssymb,amstext}
\usepackage{cite}
\usepackage{graphicx}
\usepackage{multirow}

\begin{document}

\sloppy

\rapid[Improved analytic EMRI model for eLISA data analysis]{Improved analytic extreme-mass-ratio inspiral model for scoping out eLISA data analysis}
\author{Alvin J K Chua$^1$ and Jonathan R Gair$^{1,2}$}
\address{$^1$Institute of Astronomy, University of Cambridge, Madingley Road, Cambridge CB3 0HA, United Kingdom\\$^2$School of Mathematics, University of Edinburgh, King's Buildings, Edinburgh EH9 3JZ, United Kingdom}
\ead{ajkc3@ast.cam.ac.uk}

\begin{abstract}
The space-based gravitational-wave detector eLISA has been selected as the ESA L3 mission, and the mission design will be finalised by the end of this decade. To prepare for mission formulation over the next few years, several outstanding and urgent questions in data analysis will be addressed using mock data challenges, informed by instrument measurements from the LISA Pathfinder satellite launching at the end of 2015. These data challenges will require accurate and computationally affordable waveform models for anticipated sources such as the extreme-mass-ratio inspirals (EMRIs) of stellar-mass compact objects into massive black holes. Previous data challenges have made use of the well-known analytic EMRI waveforms of Barack and Cutler, which are extremely quick to generate but dephase relative to more accurate waveforms within hours, due to their mismatched radial, polar and azimuthal frequencies. In this paper, we describe an augmented Barack--Cutler model that uses a frequency map to the correct Kerr frequencies, along with updated evolution equations and a simple fit to a more accurate model. The augmented waveforms stay in phase for months and may be generated with virtually no additional computational cost.
\end{abstract}

\pacs{04.30.Tv, 04.70.Bw, 95.30.Sf}
\submitto{\CQG}

\maketitle

\section{Introduction}
Space-based gravitational-wave (GW) astronomy will be one step closer to becoming a reality when the LISA Pathfinder mission \cite{D2015} launches at the end of 2015. A successful LISA Pathfinder flight will demonstrate the technology necessary for proposed detectors such as eLISA \cite{AEA2012} and DECIGO \cite{KEA2011} to probe the source-rich and scientifically rewarding low-frequency GW sky. The mission will also offer vital insights into the nature of the eLISA instrument; these will be used over the next few years to address several important open questions in data analysis, which must be dealt with prior to eLISA mission formulation at the end of this decade. Accurate and computationally affordable models of likely sources will allow such issues to be investigated in forthcoming mock data challenges, and are therefore urgently needed.

Extreme-mass-ratio inspirals (EMRIs)---the capture of stellar-mass compact objects (COs) by massive black holes (BHs) in galactic nuclei, such that the mass ratio of the two bodies is $\lesssim10^{-4}$---are an important type of source for space-based detectors. Radiation reaction from the emission of GWs causes the capture orbit to shrink and circularise adiabatically. During the final years of inspiral, the orbit is complicated by extreme relativistic effects as the CO is deep within the strong-field region of the BH spacetime. These effects are imprinted on the GW signal from the source; measuring them will allow us to map the multipole structure of the spacetime, and hence to test the strong-field validity of BH solutions in general relativity \cite{GVLB2013}.

The technique of matched filtering may be used to extract an EMRI signal embedded in detector noise, and to estimate the source's astrophysical parameters. Optimal detection and identification of the signal requires the generation of waveform templates from a model that is as accurate as possible. While the extreme mass ratio prohibits the use of full numerical relativity methods, it does allow the source to be modelled faithfully within the framework of BH perturbation theory. At first order in mass ratio, such models are based on the Teukolsky equation \cite{T1973}, which describes the linear perturbation to the field of a Kerr BH due to a CO moving on one of its geodesics. Orbital evolution may be introduced in Teukolsky-based models by balancing the change in orbital energy and angular momentum against the radiation fluxes. The ongoing development of gravitational self-force calculations \cite{B2009,PPV2011} will provide a more accurate evolution that accounts for both dissipative and conservative self-interaction effects at higher orders in mass ratio.

For scoping out data analysis, the large parameter space in EMRI models and the complexity of their waveforms necessitates the use of templates that can be generated as quickly as possible. As the perturbation-theory waveforms are computationally expensive, they have been supplemented in the literature by approximate waveforms designed for robust use in EMRI data analysis. The waveform model used for previous mock LISA data challenges \cite{AEA2006} is the analytic kludge (AK) \cite{BC2004}, in which the orbit is built from Keplerian ellipses, with relativistic inspiral, periapsis precession and Lense--Thirring precession imposed using analytic post-Newtonian (PN) evolution equations. The AK model is extremely quick to compute, but is less accurate than the numerical kludge (NK) model \cite{BEA2007}, which combines Kerr geodesics with PN orbital evolution for greater accuracy. In both kludge models, generation of the waveform from the orbit is sped up from perturbation-theory models by using a flat-space approximation.

In this paper, we describe an augmented AK waveform model based on a frequency mapping method. The orbital frequency and two precession rates in the AK model are matched to appropriate combinations of the fundamental frequencies for the radial, polar and azimuthal components of geodesic orbits in Kerr spacetime \cite{S2002}. We also update the AK model with suitable PN evolution equations and include an additional fit to the NK model, which itself shows excellent agreement with Teukolsky-based geodesic and inspiral waveforms \cite{H2001,DH2006}. The length of time over which the augmented AK waveform stays in phase with the NK waveform is increased by a few orders of magnitude, while the added computational cost is insignificant as the map-and-fit is only performed at the start of the orbit.

The Kerr fundamental frequencies and the parameter-space map they induce in the AK model are introduced in Sec. \ref{sec:mapping}, along with descriptions of the updated PN orbital evolution and the NK fitting method. The performance of the augmented waveform is then compared to that of the original AK waveform in Sec. \ref{sec:comparison}, with the more accurate but slower NK model used as the benchmark for both. A more detailed description and investigation of the augmented AK model will be given in a follow-up paper, while publicly available implementation code for the new waveforms will be released online shortly.

\section{Frequency mapping}\label{sec:mapping}
A bound geodesic orbit in Kerr spacetime is characterised by three fundamental frequencies $\Omega_{r,\theta,\phi}$ for the radial, polar and azimuthal components of motion. These take a simple form with the choice of a timelike parameter $\lambda=\int d\tau/\Sigma$ \cite{C1983,M2003}, where $\tau$ is proper time along the orbit and $\Sigma=r^2+a^2\cos^2\theta$ in the Boyer--Lindquist form of the Kerr metric; the frequencies are then given by \cite{DH2004,SF2015}
\begin{equation}
\Omega_r=\frac{2\pi}{\Lambda_r\Gamma},\quad\Omega_\theta=\frac{2\pi}{\Lambda_\theta\Gamma},
\end{equation}
\begin{equation}
\Omega_\phi=\lim_{N\to\infty}\frac{1}{N^2\Lambda_r\Lambda_\theta\Gamma}\int_0^{N\Lambda_r}d\lambda_r\int_0^{N\Lambda_\theta}d\lambda_\theta\;\Phi(r(\lambda_r),\theta(\lambda_\theta)),
\end{equation}
with
\begin{equation}
\Lambda_r=2\int_{r_p}^{r_a}\frac{dr}{\sqrt{R(r)}},\quad\Lambda_\theta=4\int_{\theta_\mathrm{min}}^{\pi/2}\frac{d\theta}{\sqrt{\Theta(\theta)}},
\end{equation}
\begin{equation}
\Gamma=\lim_{N\to\infty}\frac{1}{N^2\Lambda_r\Lambda_\theta}\int_0^{N\Lambda_r}d\lambda_r\int_0^{N\Lambda_\theta}d\lambda_\theta\;T(r(\lambda_r),\theta(\lambda_\theta)),
\end{equation}
where $R(r)$, $\Theta(\theta)$, $\Phi(r,\theta)$ and $T(r,\theta)$ are the usual potential functions in the MTW form of the Kerr geodesic equations \cite{MTW1973}. The orbit is specified by the set of parameters $(r_p,r_a,\theta_\mathrm{min})$ (the values of $r$ at periapsis and apoapsis, and the minimal value of $\theta$ respectively), which fully describes the range of motion in the radial and polar coordinates. Expressions for $\Omega_{r,\theta,\phi}$ in terms of the alternative parametrisation $(e,\iota,p)$ (the quasi-Keplerian eccentricity, inclination and semi-latus rectum of the orbit respectively) have also been given by Schmidt \cite{S2002}.

In the Newtonian limit, the fundamental frequencies reduce to a single orbital frequency $f_\mathrm{orb}=\omega_r=\omega_\theta=\omega_\phi$, where we have defined the non-angular and dimensionful frequencies
\begin{equation}
\omega_{r,\theta,\phi}:=\frac{\Omega_{r,\theta,\phi}}{2\pi M}
\end{equation}
for a BH of mass $M$. The frequency $f_\mathrm{orb}$ appears as $\nu$ in the AK model of Barack and Cutler \cite{BC2004}, where it is related to $e$ and $p$ by Kepler's third law. Precession of the orbital ellipse about the orbital angular momentum vector and precession of the orbital plane about the BH spin vector are introduced and evolved separately in the model; these are described by the periapsis precession frequency $f_\mathrm{peri}$ and the Lense--Thirring frequency $f_\mathrm{LT}$ respectively. A representative GW frequency may be defined as twice the azimuthal orbital frequency (i.e. $f_\mathrm{GW}:=2(f_\mathrm{orb}+f_\mathrm{peri})$), which is the dominant GW harmonic in the circular case $e=0$.$^1$\footnote[0]{$^1$In the parametrised notation of Barack and Cutler \cite{BC2004}, we have $f_\mathrm{orb}=\dot{\Phi}/(2\pi)$, $f_\mathrm{peri}=(\dot{\tilde{\gamma}}+\dot{\alpha})/(2\pi)$ and $f_\mathrm{LT}=\dot{\alpha}/(2\pi)$. Our representative GW frequency differs slightly from their choice $f_2=(\dot{\Phi}+\dot{\tilde{\gamma}})/\pi$ for the dominant $n=2$ harmonic.}

For any given point $(e,\iota,p)$ along the AK inspiral trajectory, the orbital, periapsis and Lense--Thirring frequencies are determined by Kepler's third law and 1.5PN expressions for the two precession rates \cite{BC2004}. However, if we are to interpret $(e,\iota,p)$ as the quasi-Keplerian parameters of a Kerr geodesic, $f_\mathrm{orb}$, $f_\mathrm{peri}$ and $f_\mathrm{LT}$ will not in general equal the correct values $\omega_r$, $\omega_\phi-\omega_r$ and $\omega_\phi-\omega_\theta$ respectively. Matching the three frequencies in the AK model with the appropriate combinations of $\omega_{r,\theta,\phi}$ then induces a three-dimensional endomorphism over the AK parameter space; we choose to map the BH mass $M$, the BH spin parameter $a$ and the semi-latus rectum $p$ to some unphysical values $\tilde{M}$, $\tilde{a}$ and $\tilde{p}$.$^2$\footnote[0]{$^2$Serendipitous acronymisation aside, our choice of parameters gives better results than, say, the map $(e,\iota,p)\mapsto(\tilde{e},\tilde{\iota},\tilde{p})$. This is because periapsis precession and Lense--Thirring precession are more directly determined by the central mass $M$ and its rotation $a$ respectively.} The map $(M,a,p)\mapsto(\tilde{M},\tilde{a},\tilde{p})$ is given implicitly by solving the algebraic system of equations
\begin{equation}
f_\mathrm{orb}(\tilde{M},\tilde{a},\tilde{p})=\omega_r(M,a,p),
\end{equation}
\begin{equation}
f_\mathrm{peri}(\tilde{M},\tilde{a},\tilde{p})=\omega_\phi(M,a,p)-\omega_r(M,a,p),
\end{equation}
\begin{equation}
f_\mathrm{LT}(\tilde{M},\tilde{a},\tilde{p})=\omega_\phi(M,a,p)-\omega_\theta(M,a,p)
\end{equation}
for $(\tilde{M},\tilde{a},\tilde{p})$, which we define as the root closest to the physical parameters $(M,a,p)$ with a Euclidean metric on parameter space.$^3$\footnote[0]{$^3$In practice, the closest root is usually obvious for any sensible choice of metric.}

Substituting the unphysical parameters $(\tilde{M},\tilde{a},\tilde{p})$ for $(M,a,p)$ in the AK model provides an instantaneous correction of the frequencies at any point along the inspiral trajectory. In order to keep the added computational cost as low as possible, we only evaluate the map at a single point $(e_t,\iota_t,p_t)$ on the physical trajectory; using the mapped values $(\tilde{M},\tilde{a},\tilde{p}_t)$, we then evolve $(e,\tilde{p})$ with higher-order 3PN $\mathcal{O}(e^6)$ expressions given by Sago and Fujita \cite{SF2015}, leaving the inclination $\iota$ constant as in the original AK approximation (where it appears as $\lambda$). We also evolve $f_\mathrm{orb}$, $f_\mathrm{peri}$ and $f_\mathrm{LT}$ with the appropriate combinations of 3PN $\mathcal{O}(e^6)$ expressions for the fundamental frequencies \cite{SF2015}, such that they remain consistent with the orbital evolution.

Information from more accurate EMRI models may also be incorporated. We further augment the mapped AK model by fitting the inspiral trajectory to that in the NK model of Babak et al. \cite{BEA2007}, which uses Teukolsky-fitted mixed-order PN expressions \cite{GG2006} to evolve its orbits. Evaluation of the map at each point along the NK trajectory gives a ``best-fit'' trajectory of the AK parameters $(\tilde{M},\tilde{a},e,\tilde{p})_\mathrm{fit}$, where the unphysical BH mass and spin parameters now evolve along the inspiral as well. A local quadratic fit to the best-fit trajectory is implemented by computing only three consecutive points on the NK trajectory and the map at each point (i.e. the evolution of $(\tilde{M},\tilde{a},e,\tilde{p})_\mathrm{fit}-(\tilde{M},\tilde{a},e,\tilde{p})$ is taken to be quadratic in time, with its coefficients given by finite difference quotients obtained from the three points). This increases the computational cost by a fixed but essentially insignificant amount.

\section{Waveform comparison}\label{sec:comparison}

\begin{figure}
\centering
\includegraphics[width=0.8\columnwidth]{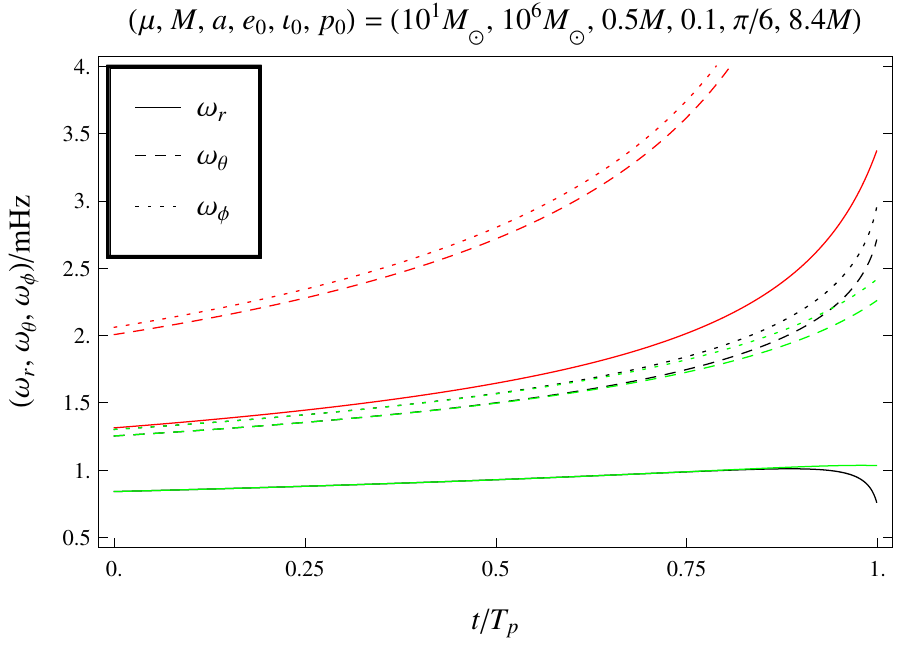}
\caption{Evolution of fundamental frequencies $\omega_{r,\theta,\phi}$ in original AK model (red), augmented AK model (green) and NK model (black), for generic EMRI plunging at $t=T_p=1\,\mathrm{yr}$.}
\label{fig:frequencies}
\end{figure}

We now consider the map-and-fit described in Sec. \ref{sec:mapping} for a generic EMRI with a CO mass of $\mu=10\,M_\odot$, a BH mass of $M=10^6\,M_\odot$, a BH spin of $a=0.5M$, an initial eccentricity of $e_0=0.1$ and an initial inclination of $\iota_0=\pi/6$. An initial semi-latus rectum of $p_0=8.4M$ is chosen such that the CO plunges (reaches the last stable orbit of the BH) exactly one year after entering the eLISA band at $f_\mathrm{GW}=2.6\,\mathrm{mHz}$. Fig. \ref{fig:frequencies} shows a plot of the fundamental frequencies $\omega_{r,\theta,\phi}$ in the original AK, augmented AK and NK models; it is clear that the accumulated phase error in the AK model (with respect to the NK model) is drastically reduced by our frequency mapping method.

Kludge waveforms are generated with a flat-space multipole formula, which offers significant computational savings over perturbation-theory waveforms. For the augmented AK model, we retain the analytic mode-sum approximation of Peters and Mathews \cite{PM1963} used in the original model. This quadrupolar waveform is extremely quick to compute for low-eccentricity orbits, although the number of required modes scales linearly with $e$ \cite{BC2004}. We use a quadrupolar (but numerically integrated) NK waveform as well; more accurate versions of the NK model have been implemented using formulae for higher-order moments or fast-motion sources \cite{BEA2007}, but these do not show sufficient improvement in accuracy to warrant the additional computational expense in the current work. Both AK models are quicker than the NK model by a factor of two (for $e_0\approx0.5$) to 10 (for $e_0\approx0.1$) in the case of year-long waveforms sampled every $30\,\mathrm{s}$, with a greater speed-up if longer waveforms are considered.

\begin{figure}
\centering
\includegraphics[width=0.8\columnwidth]{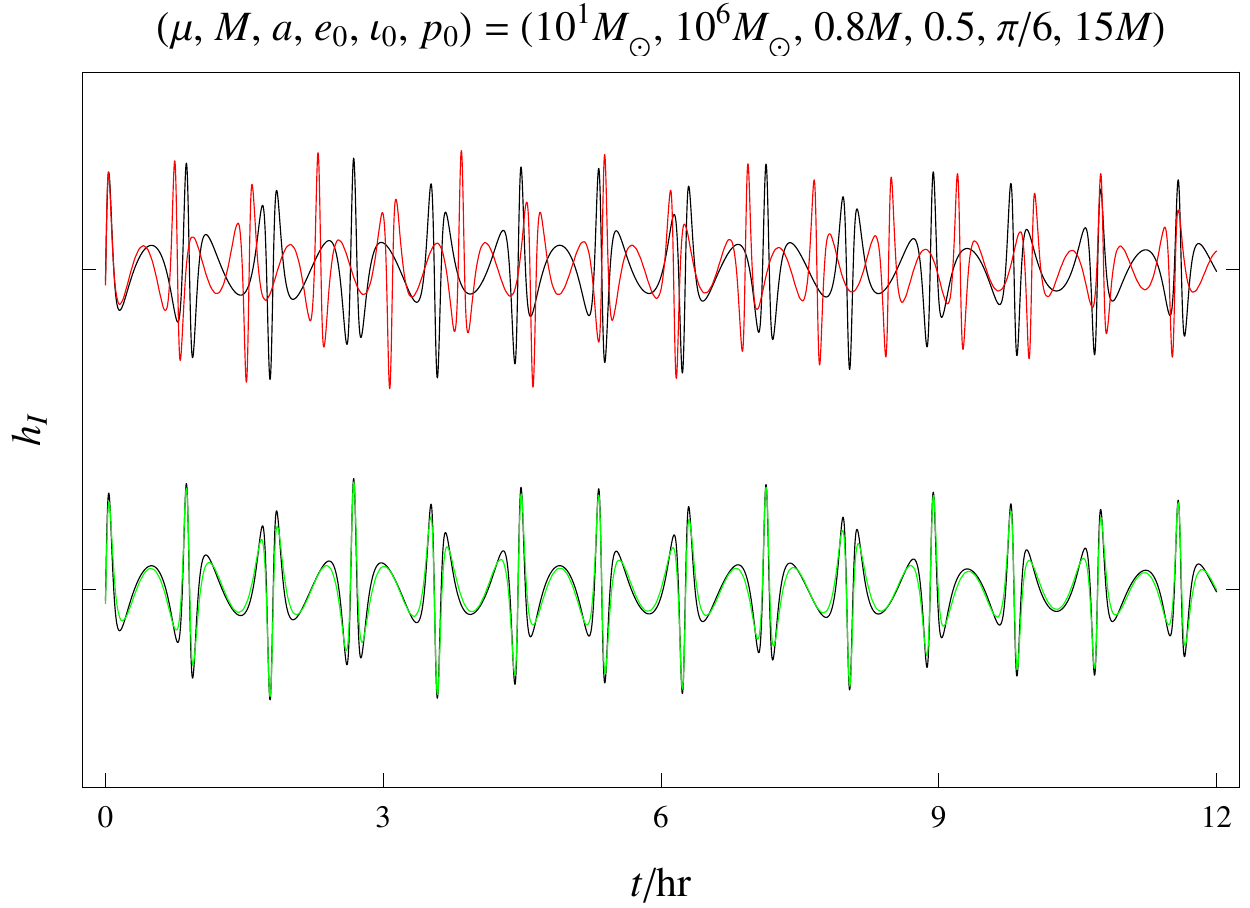}
\caption{First 12 hours of original (red) and augmented (green) AK waveforms overlaid on NK waveform (black), for generic EMRI with orbital parameters $(e,\iota,p)=(0.5,\pi/6,15M)$ at $t=0$.}
\label{fig:waveforms}
\end{figure}

Even at the early-inspiral stage, the original AK waveform is known to dephase rapidly with respect to the NK waveform. This is illustrated in the top plot of Fig. \ref{fig:waveforms}, which shows both waveforms (more precisely, the eLISA response functions $h_I(t)$ \cite{BC2004}) for another generic EMRI with an initial semi-latus rectum of $p_0=15M$; while the AK waveform does capture the main features of the NK waveform, it is a full cycle out of phase within three hours. The severe dephasing is due to the mismatched fundamental frequencies in the two models, and hence is well mitigated by our frequency mapping method (as seen from the bottom plot of Fig. \ref{fig:waveforms}).

To compare the accuracy of the original and augmented AK waveforms over longer timescales, it is useful to introduce the standard matched-filtering inner product (on the space of finite-length time series) between two waveforms $a(t)$ and $b(t)$, i.e. \cite{CF1994}
\begin{equation}
\langle a|b\rangle=2\int_0^\infty df\;\frac{\tilde{a}^*(f)\tilde{b}(f)+\tilde{a}(f)\tilde{b}^*(f)}{S_n(f)},
\end{equation}
where $S_n(f)$ is the eLISA noise power spectral density (for which we use an analytic approximation \cite{AEA2013}). A measure of the accuracy with which $a$ represents $b$ (or vice versa) is then given by the overlap function
\begin{equation}\label{eq:overlap}
\mathcal{O}(a|b):=\frac{\langle a|b\rangle}{\sqrt{\langle a|a\rangle\langle b|b\rangle}},
\end{equation}
which takes the value of one for identical waveforms and zero for orthogonal waveforms.

We use \eref{eq:overlap} to quantify the accuracy of the waveform $h(t)$ in both the original and augmented AK models, with respect to the NK waveform.$^4$\footnote[0]{$^4$Both detector channels are considered for each model, i.e. $h(t):=h_I(t)+ih_{II}(t)$.} The first row of Tab. \ref{tab:table} gives the overlap $\mathcal{O}(h_\mathrm{AK}|h_\mathrm{NK})_T$ over different timescales $T$ for the EMRI considered in Fig. \ref{fig:frequencies}, along with the average ratio of computation times $\tau_\mathrm{NK}/\tau_\mathrm{AK}$. Also given in Tab. \ref{tab:table} are the corresponding values for the same source with (i) a lower CO mass, (ii) a higher BH spin and (iii) a higher initial eccentricity. For each EMRI, $p_0$ is chosen such that the CO plunges at $t=1\,\mathrm{yr}$. Our frequency mapping method improves the accuracy of the AK waveform by a factor of $\gtrsim100$ for the source in Fig. \ref{fig:frequencies}, with the overlap over the first two months approaching the accuracy required for actual parameter estimation. While the two-month overlap is still poor for lower mass ratios (since the orbit begins further into the strong field) and higher eccentricity (due to the richer structure of the waveforms), the length of time over which the AK waveform remains phase-coherent with the NK waveform is increased from under an hour to over two months for all four sources.

\begin{table}
\caption{\label{tab:table}Average ratio of computation times and waveform overlaps over six/two months, for generic EMRIs with different CO mass/BH spin/initial eccentricity.}
\begin{tabular}{@{}lllllll}
\br
\multirow{2}{*}{$\left(\frac{\mu}{M_\odot},\frac{a}{M},e_0,\frac{p_0}{M}\right)$}&\multirow{2}{*}{$\frac{\tau_\mathrm{NK}}{\tau_\mathrm{AK}}$}&\multicolumn{2}{c}{$\mathcal{O}(h_\mathrm{AK}|h_\mathrm{NK})_{6\,\mathrm{mth}}$}&\multicolumn{2}{c}{$\mathcal{O}(h_\mathrm{AK}|h_\mathrm{NK})_{2\,\mathrm{mth}}$}\\
\cline{3-6}\ms
&&Orig.&Augm.&Orig.&Augm.\\
\ns\mr
$(10^1,0.5,0.1,8.4)$&$9.5$&$6.4\times10^{-4}$&$1.5\times10^{-1}$&$2.0\times10^{-3}$&$9.5\times10^{-1}$\\
$(10^0,0.5,0.1,5.9)$&$13$&$2.7\times10^{-6}$&$1.3\times10^{-1}$&$4.4\times10^{-5}$&$5.9\times10^{-1}$\\
$(10^1,0.8,0.1,7.7)$&$11$&$-2.7\times10^{-4}$&$1.3\times10^{-1}$&$-1.6\times10^{-3}$&$9.3\times10^{-1}$\\
$(10^1,0.5,0.5,8.2)$&$1.6$&$-2.0\times10^{-3}$&$4.3\times10^{-2}$&$8.6\times10^{-3}$&$2.3\times10^{-1}$\\
\br
\end{tabular}
\end{table}

\section{Conclusion}
The well-known AK waveform model of Barack and Cutler \cite{BC2004} is extremely quick to generate, but dephases within hours with respect to more accurate EMRI waveforms due to a mismatch of its radial, polar and azimuthal frequencies with the actual Kerr frequencies along its orbit. Hence using AK waveforms as search templates for EMRIs will result in reduced signal-to-noise ratios for detection, as well as an inaccurate estimation of astrophysical parameters. This limits their utility in scoping out data analysis issues for space-based GW detectors---an area where there are several important open questions that must be addressed within the next few years, in preparation for eLISA mission formulation at the end of this decade.

We have proposed in this paper an augmented AK waveform model that features a frequency map to the Kerr frequencies, updated PN evolution equations and a quadratic fit to the NK model of Babak et al. \cite{BEA2007}. This new waveform is virtually as quick to compute as its predecessor, but stays in phase (with respect to the NK waveform) for months. Our model is a useful addition to the inventory of EMRI waveforms for the analysis of data from space-based GW detectors, as well as an essential tool for near-future work on eLISA data analysis and mock eLISA data challenges; it can also be upgraded with higher-order fits to improved models (e.g. an NK waveform including self-force evolution \cite{GEA2011,WEA2012}) when these become available.

\section*{Acknowledgements}
We thank Christopher Moore for helpful discussions. AJKC's work was supported by the Cambridge Commonwealth, European and International Trust. JRG's work was supported by the Royal Society.

\section*{References}
\bibliographystyle{unsrt}
\bibliography{references}

\end{document}